\documentclass[onecolumn,preprintnumbers,amsmath,amssymb]{revtex4}

\usepackage{float}
\usepackage{graphicx}
\usepackage{dcolumn}
\usepackage{bm}
\usepackage[normalem]{ulem}
\newcommand{\be}{\begin{equation}}
\newcommand{\ee}{\end{equation}}
\newcommand{\bea}{\begin{eqnarray}}
\newcommand{\eea}{\end{eqnarray}}
\newcommand{\beaa}{\begin{eqnarray*}}
\newcommand{\eeaa}{\end{eqnarray*}}

\begin{document}

\preprint{}

\title{Decoherence-free subspace and entanglement sudden death of multi-photon polarization states in fiber channels}
\author{Yiwen Liu$^1$}%
\affiliation{%
$^1$School of Physics, Peking University, Beijing, China}%

\date{\today}

\begin{abstract}
The construction of quantum networks requires long-distance teleportation of multi-qubit entangled states. Here, we investigate the entanglement dynamics of GHZ and W states in fiber channels. In a fiber channel, the two most important phenomena that affect polarization entanglement are polarization mode dispersion (PMD) and polarization-dependent loss (PDL). We theoretically characterize how PMD and PDL vectors affect three-qubit states. In particular, upon quantifying the entanglement at the output states using concurrence and entanglement witness, we reveal the occurrence of entanglement sudden death (ESD) and the appearance of decoherence-free subspaces (DSFs) in tripartite systems. Finally, we explore the evolution of GHZ and W state with an arbitrary number of photons in a fiber network and evaluate the expectation value of the entanglement witness.
\end{abstract}

\maketitle

\section{Introduction}
Quantum entanglement is a crucial ingredient for quantum teleportation, quantum key distribution, and quantum computing. Due to the current demand for encrypted communication, quantum teleportation has emerged as one of the core topics in the field of quantum information \cite{Chen2021,Yin2017}. Among the various areas related to entanglement, multi-qubit entanglement and decoherence mechanisms in the quantum channels are two relevant topics for the development of quantum networks.

By introducing multi-partite entanglement systems, novel quantum protocols allowing for more applications than two-partite systems can be implemented\cite{Zoller2005}, such as secret sharing or multi-partite fingerprinting. Therefore, a great deal of work has been devoted to analyzing entanglement for an increasing number of particles. Relying on the development of spontaneous parametric down-conversion and multi-photon interferometry, the number of photons entangled simultaneously has grown from 8 in 2012\cite{Yao2012}, to 10 in 2016\cite{Wang2016}, and to 12 in 2018\cite{Zhong2019}. In addition, attention has been paid to how multiparticle systems interact with noisy environments\cite{Siomau2010}. Weinstein explored the dynamics of tripartite entanglement under dephasing via concurrence and tripartite negativity\cite{Weinstein2009}; Zhang investigated GHZ and W states evolution in depolarizing and the amplitude-damping channels via negativity and $\pi$-tangle\cite{Zhang2011}. These results have contributed to understanding the evolution of complex quantum systems.

Further research has been devoted to analyzing the decoherence of entangled states in real fiber channels. Riccardi developed an effective model describing the combined effect of dephasing and mode-filtering elements on Bell states in a single channel, which has been experimentally verified \cite{Riccardi2021}. Kozubov discussed the use of the Liouville equation to analyze the dynamics of a single-photon state and estimated quantum bit error rates as a function of the fiber length\cite{Kozubov2019}. Most of these studies involve only two-photon entangled states, with only a few considering different optical channels and noise sources. Therefore, the behavior of multipartite entanglement in fiber channels is still an unsettled area to a considerable degree.

For nonlocal entangled systems, previous works have reported two special decoherence phenomena: entanglement sudden death (ESD) and decoherence-free subspace(DSF). The former refers to situations in which the entanglement vanishes in a finite time, whereas DSF refers to situations in which entangled states entirely contained in a subspace of the entire Hilbert state are immune to decoherence. The ESD and DSF phenomena in bipartite systems have been exhaustively explored and observed experimentally\cite{Antonelli2011,Kwiat2000}. This is not the case for multi-particle systems, where ESD and DSF have not been thoroughly studied. 

In this paper, we present a quantitative analytical model that describes the decoherence of multiphoton polarization-entangled states in single-mode fibers taking into account polarization mode dispersion(PMD) and polarization-dependent loss(PDL), and shows the DSF and ESD that may occur in quantum network.

The paper is organized as follows. At first, we analyze the dynamics of three-qubit GHZ and W states propagating in three separate fibers and subject to independent PMD or PDL. Due to the randomness of PMD and PDL, we use two parameters, the differential group delay $\tau$ and the loss coefficient $\gamma$ , to describe the evolution of the density matrix. We will use two different entanglement criteria – concurrence and entanglement witness-- to quantify and reveal bipartite and tripartite entanglement and to investigate the appearance of DSF and the occurrence of ESD in three-photon systems. Finally, we extend the model to any number of entangled photons and calculate the expectation value of the entanglement witness after propagation in fibers.

\section{Decoherence of GHZ and W states in fiber channels}
In this Section, we explore the decoherence of three-photon entangled states passing through PMD and PDL elements in a quantum network.

\subsection{Decoherence induced by PMD and PDL}
According to previous studies, among the optical effects occurring in fiber, such as chromatic dispersion or fiber nonlinearity, the main source of noise in single-mode fiber channels originates from two optical effects - polarization mode dispersion caused by the birefringence and polarization-dependent loss caused by the asymmetric structure of the fiber\cite{Riccardi2021}. 
The PMD effect decompose the input pulse into a pair of orthogonal polarization states, which are delayed relative to each other. And the PDL element in the fiber represents a typical example of a mode-filtering operation, which relatively increases the intensity of one mode in a superposition state over the other. The rotation and intensity change of the polarization modes caused by PMD and PDL can be described by the transmission matrix. The transmission matrix corresponding to PMD in the first-order approximation can be expressed as\cite{Huttner2000}: 
\bea U\left( \omega  \right)\text{=}\exp \left( \frac{-i\omega \vec{\tau }\cdot \vec{\sigma }}{2} \right) \label{1}\eea

The orientation of the PMD vector $\vec{\tau}$  corresponds to the principal states of polarization(PSP), which are defined as the slowest and fastest eigenstates. The magnitude of $\vec{\tau}$  is equal to the differential group delay(DGD) $\tau=|\vec{\tau}|$ \cite{Ruan2021};On the other hand, the transmission matrix for PDL effects may be expressed as\cite{Huttner2000}:
\bea T=\exp \left( \frac{\vec{\gamma }\cdot \vec{\sigma }}{2} \right) \label{2}\eea
where the orientation of the PDL vector $\vec{\gamma}$ corresponds to polarization modes with the maximum and minimum loss. The magnitude of $\vec{\gamma}$ is equal to the loss coefficient $\gamma=|\vec{\gamma}|$ . PMD and PDL lead to two different types of decoherence processes, so we need to separately discuss the evolution of entanglement for states subject to the two different kinds of noise.

\subsection{Concurrence and Entanglement Witness for three-qubit states}
For more than two qubits, maximally entangled states have two possible forms—GHZ(Greenberger-Horne-Zeilinger) state and W state. We assume that a three-photon maximally entangled state at the input of the fiber is prepared as GHZ state or W state:
\bea 
{{\left| \psi  \right\rangle }_{GHZ}}\text{=}\frac{1}{\sqrt{2}}\left( \left| 000 \right\rangle \text{+}\left| 111 \right\rangle  \right),
{{\left| \psi  \right\rangle }_{W}}\text{=}\frac{1}{\sqrt{3}}\left( \left| 001 \right\rangle \text{+}\left| 010 \right\rangle +\left| 100 \right\rangle  \right) 
\label{3}\eea
where $|0\rangle$   and $|1\rangle$  are the orthogonal polarization bases for the three photons A, B, and C. 

Next, we discuss how to quantify the entanglement at the output of fiber channels. The concurrence is a sufficient measure for bipartite entanglement. After partial trace over one qubit , the concurrence between the remaining qubits i and j can be defined as ${{C}_{ij}}=\max \left\{ 0,\sqrt{{{\lambda }_{1}}}-\sqrt{{{\lambda }_{2}}}-\sqrt{{{\lambda }_{3}}}-\sqrt{{{\lambda }_{4}}} \right\}$, where ${{\lambda }_{1}}$,${{\lambda }_{2}}$, ${{\lambda }_{3}}$, ${{\lambda }_{4}}$ are the eigenvalues of the matrix $\varsigma \text{=}{{\rho }_{ij}}\left( \sigma _{y}^{i}\otimes \sigma _{y}^{j} \right){{\rho }_{ij}}^{*}\left( \sigma _{y}^{i}\otimes \sigma _{y}^{j} \right)$. ${{\rho }_{ij}}$ denotes the reduced density matrix for qubits i and j, and ${{\sigma }_{y}}^{i}$ is the Pauli matrix of the qubit i. The concurrence between any two qubits is zero for GHZ state ${{C}_{AB}}={{C}_{AC}}={{C}_{BC}}=0$, whereas the concurrence for W state is given ${{C}_{AB}}={{C}_{AC}}={{C}_{BC}}=\frac{2}{3}$\cite{Weinstein2009}. 

However, so far, there is still no general analytical method to calculate the concurrence of the tripartite mixed state\cite{Cao2010}. Alternatively, we can use the entanglement witness to measure tripartite entanglement. The entanglement witness is a Hermitian observable which can efficiently detect entanglement experimentally. Given an output state for a quantum channel, we can construct one or even more forms of entanglement witnesses, with non-negative expectation values for separable states and negative values for entangled states.

A widely used entanglement witness can be constructed as follows\cite{Riccardi2021a}:
\bea EW={{a}_{0}}I-\left| \psi  \right\rangle \left\langle  \psi  \right|,{{a}_{0}}=\underset{\left| \varphi  \right\rangle \in P}{\mathop{\max }}\,{{\left| \left\langle \varphi \left| \psi  \right\rangle  \right. \right|}^{2}}    \label{4}\eea
where $P$ represents the set of all possible bipartite states $\left| \varphi  \right\rangle $. $\left| \psi  \right\rangle$ refers to the initial pure entangled state, i.e. GHZ state or W state in our case.

The expectation value, denoted by $V$, of the operator $EW$ is given by 
\bea  V=Tr\left( EW\rho  \right)={{a}_{0}}-F={{a}_{0}}-\left\langle  \psi  \right|\rho \left| \psi  \right\rangle   \label{5}\eea
where $F$ denotes the fidelity of the mixed state $\rho$ to the state $\left| \psi  \right\rangle$. The values of $V$ for GHZ state and W state are different. For a GHZ state we have:
\bea {{a}_{0}}=\frac{1}{2},{{V}_{GHZ}}=\frac{1}{2}-\left\langle  {{\psi }_{GHZ}} \right|\rho \left| {{\psi }_{GHZ}} \right\rangle, V_{GHZ}^{0}=-\frac{1}{2}    \label{6}\eea
whereas for a W state:
\bea {{a}_{0}}=\frac{2}{3},{{V}_{W}}=\frac{2}{3}-\left\langle  {{\psi }_{W}} \right|\rho \left| {{\psi }_{W}} \right\rangle, V_{W}^{0}=-\frac{1}{3}    \label{7}\eea

$V_{GHZ}^{0}$ and $V_{W}^{0}$ are the expectation values for pure GHZ state and W state. In summary, by using concurrence and entanglement witnesses, we can characterize bipartite and tripartite entanglement in a three-qubit system.

\subsection{Decoherence induced by PMD effects for GHZ and W states}
Evaluating the decoherence occurring in fibers is challenging since the orientation and magnitude of the PMD and PDL vectors change randomly in different channels\cite{Antonelli2011,Kirby2018}. Specifically, previous studies have shown that when the PSPs in two fibers are aligned with the polarization basis, Bell states may get maximum achievable concurrence \cite{Ruan2021}. To simplify the calculation, we assume that the PMD and the PDL vectors in our fiber channels are both aligned with the initial polarization basis of GHZ and W states, which can be achieved by performing the local rotation operation on each photon. 

We consider the schematic quantum network illustrated in Fig.1, in which three entangled photons are propagating in optical fibers with PMD:

\begin{figure}[H]
	\centering
		\includegraphics[width=.75\columnwidth]{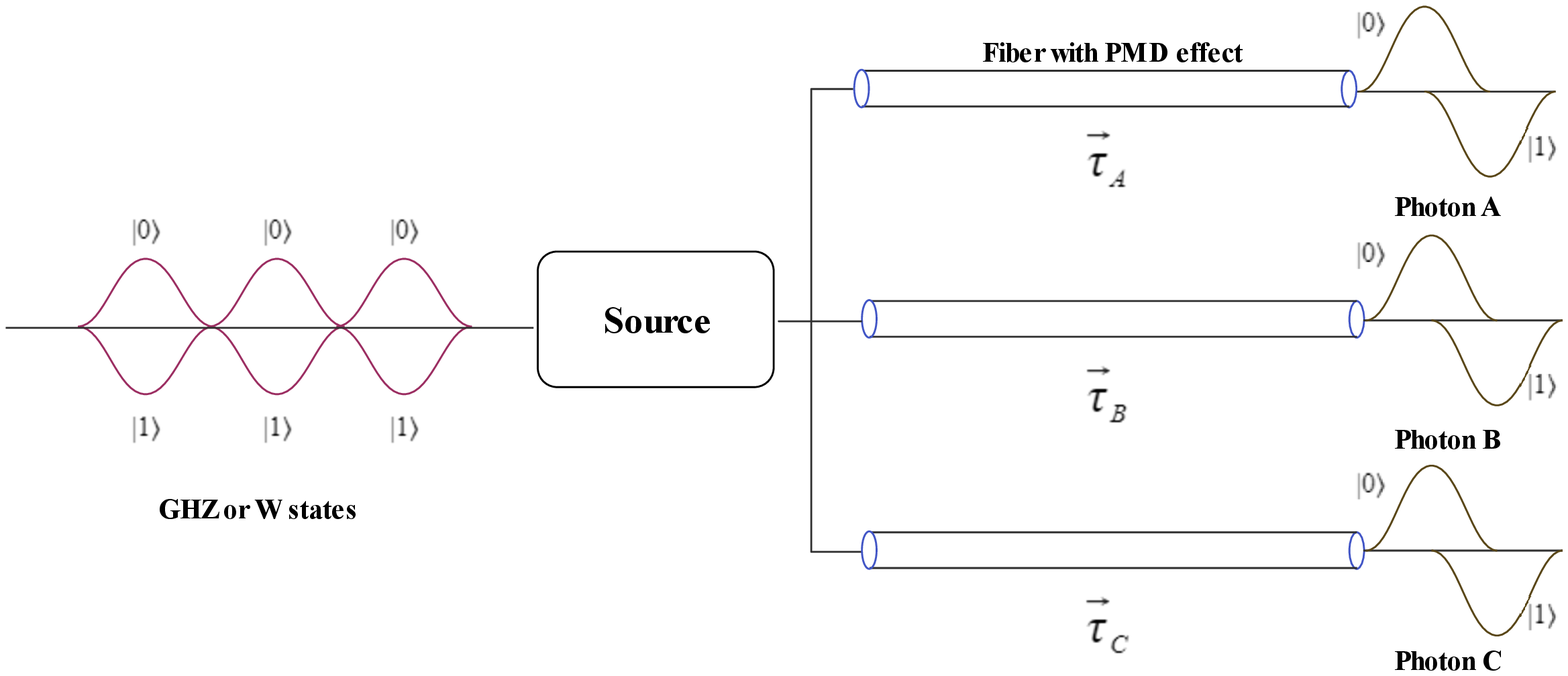}
		\caption{A quantum network, where PMD in different fibers introduces different time delays ${{\tau }_{A}}$,${{\tau }_{B}}$,${{\tau }_{C}}$ in the polarization states, assuming that the polarization bases of each photon, $\left| 0 \right\rangle $ and $\left| 1 \right\rangle $, are always aligned with the slow and fast PSP in the channels. In this case, the polarization bases become the eigenstates of the operator $U\left( \omega  \right)$:$U\left( \omega  \right)\left| 0 \right\rangle \text{=}\exp \left( \frac{-i\omega \tau }{2} \right)\left| 0 \right\rangle ,U\left( \omega  \right)\left| 1 \right\rangle \text{=}\exp \left( \frac{i\omega \tau }{2} \right)\left| 1 \right\rangle $.} \label{Figure1}
\end{figure}

By applying the transmission matrix $U\left( \omega  \right)$ to GHZ and W state, we can derive the output polarization state as follows:

\begin{footnotesize}
\bea    
\begin{aligned}
	 \psi _{GHZ}^{1}&=\exp \left( \frac{-i{{\omega }_{A}}{{{\vec{\tau }}}_{A}}\cdot \vec{\sigma }}{2} \right)\exp \left( \frac{-i{{\omega }_{B}}{{{\vec{\tau }}}_{B}}\cdot \vec{\sigma }}{2} \right)\exp \left( \frac{-i{{\omega }_{C}}{{{\vec{\tau }}}_{C}}\cdot \vec{\sigma }}{2} \right){{\psi }_{GHZ}} \\ 
	& =\iint{d{{\omega }_{A}}}d{{\omega }_{B}}d{{\omega }_{C}}\tilde{f}\left( {{\omega }_{A}},{{\omega }_{B}},{{\omega }_{C}} \right)\left| {{\omega }_{A}},{{\omega }_{B}},{{\omega }_{C}} \right\rangle  \\ 
	& \otimes \frac{1}{\sqrt{2}}\left[ \exp \left( \frac{-i{{\omega }_{A}}{{\tau }_{A}}-i{{\omega }_{B}}{{\tau }_{B}}-i{{\omega }_{C}}{{\tau }_{C}}}{2} \right)\left| 000 \right\rangle \text{+}\exp \left( \frac{i{{\omega }_{A}}{{\tau }_{A}}+i{{\omega }_{B}}{{\tau }_{B}}+i{{\omega }_{C}}{{\tau }_{C}}}{2} \right)\left| 111 \right\rangle  \right], \\ 
	 \psi _{W}^{1}&=\exp \left( \frac{-i{{\omega }_{A}}{{{\vec{\tau }}}_{A}}\cdot \vec{\sigma }}{2} \right)\exp \left( \frac{-i{{\omega }_{B}}{{{\vec{\tau }}}_{B}}\cdot \vec{\sigma }}{2} \right)\exp \left( \frac{-i{{\omega }_{C}}{{{\vec{\tau }}}_{C}}\cdot \vec{\sigma }}{2} \right){{\psi }_{W}} \\ 
	& =\iint{d{{\omega }_{A}}}d{{\omega }_{B}}d{{\omega }_{C}}\tilde{f}\left( {{\omega }_{A}},{{\omega }_{B}},{{\omega }_{C}} \right)\left| {{\omega }_{A}},{{\omega }_{B}},{{\omega }_{C}} \right\rangle  \\ 
	& \otimes \frac{1}{\sqrt{3}}\left[ \exp \left( \frac{-i{{\omega }_{A}}{{\tau }_{A}}-i{{\omega }_{B}}{{\tau }_{B}}\text{+}i{{\omega }_{C}}{{\tau }_{C}}}{2} \right)\left| 001 \right\rangle \text{+}\exp \left( \frac{-i{{\omega }_{A}}{{\tau }_{A}}+i{{\omega }_{B}}{{\tau }_{B}}-i{{\omega }_{C}}{{\tau }_{C}}}{2} \right)\left| 010 \right\rangle +\exp \left( \frac{i{{\omega }_{A}}{{\tau }_{A}}-i{{\omega }_{B}}{{\tau }_{B}}-i{{\omega }_{C}}{{\tau }_{C}}}{2} \right)\left| 100 \right\rangle  \right]  
\end{aligned}
\label{8}\eea
\end{footnotesize}
where ${{\tau }_{A}}$,${{\tau }_{B}}$,${{\tau }_{C}}$ denote the differential group delays in different channels, and $\tilde{f}\left( {{\omega }_{A}},{{\omega }_{B}},{{\omega }_{C}} \right)$ represents the frequency content of the entangled photons. The density matrix of GHZ and W states after passing through PMD can be obtained by tracing over the frequency modes:
\bea   
\begin{aligned}
&{{\rho }_{GHZ}}=\frac{1}{2}\left( \begin{matrix}
	1 & 0 & 0 & 0 & 0 & 0 & 0 & {{R}^{*}}\left( {{\tau }_{A}},{{\tau }_{B}},{{\tau }_{C}} \right)  \\
	0 & 0 & 0 & 0 & 0 & 0 & 0 & 0  \\
	0 & 0 & 0 & 0 & 0 & 0 & 0 & 0  \\
	0 & 0 & 0 & 0 & 0 & 0 & 0 & 0  \\
	0 & 0 & 0 & 0 & 0 & 0 & 0 & 0  \\
	0 & 0 & 0 & 0 & 0 & 0 & 0 & 0  \\
	0 & 0 & 0 & 0 & 0 & 0 & 0 & 0  \\
	R\left( {{\tau }_{A}},{{\tau }_{B}},{{\tau }_{C}} \right) & 0 & 0 & 0 & 0 & 0 & 0 & 1  \\
\end{matrix} \right),\\
&{{\rho }_{W}}=\frac{1}{3}\left( \begin{matrix}
	0 & 0 & 0 & 0 & 0 & 0 & 0 & 0  \\
	0 & 1 & {{R}^{*}}\left( 0,{{\tau }_{B}},-{{\tau }_{C}} \right) & 0 & {{R}^{*}}\left( {{\tau }_{A}},0,-{{\tau }_{C}} \right) & 0 & 0 & 0  \\
	0 & R\left( 0,{{\tau }_{B}},-{{\tau }_{C}} \right) & 1 & 0 & {{R}^{*}}\left( {{\tau }_{A}},-{{\tau }_{B}},0 \right) & 0 & 0 & 0  \\
	0 & 0 & 0 & 0 & 0 & 0 & 0 & 0  \\
	0 & R\left( {{\tau }_{A}},0,-{{\tau }_{C}} \right) & R\left( {{\tau }_{A}},-{{\tau }_{B}},0 \right) & 0 & 1 & 0 & 0 & 0  \\
	0 & 0 & 0 & 0 & 0 & 0 & 0 & 0  \\
	0 & 0 & 0 & 0 & 0 & 0 & 0 & 0  \\
	0 & 0 & 0 & 0 & 0 & 0 & 0 & 0  \\
\end{matrix} \right)\\
\end{aligned}
\label{9}\eea

For three qubits with uncorrelated frequency distributions, the function $R\left( {{\tau }_{A}},{{\tau }_{B}},{{\tau }_{C}} \right)$ is obtained by \cite{Shtaif2011}:
\bea    
\begin{aligned}
	& R\left( {{\tau }_{A}},{{\tau }_{B}},{{\tau }_{C}} \right)={{\iiint{\left| \tilde{f}\left( {{\omega }_{A}},{{\omega }_{B}},{{\omega }_{C}} \right) \right|}}^{2}}{{e}^{i\left( {{\omega }_{A}}{{\tau }_{A}}+{{\omega }_{B}}{{\tau }_{B}}+{{\omega }_{C}}{{\tau }_{C}} \right)}}d{{\omega }_{A}}d{{\omega }_{B}}d{{\omega }_{C}} \\ 
	& \text{=}{{\iiint{\left| f\left( {{\omega }_{A}} \right) \right|}}^{2}}{{\left| f\left( {{\omega }_{B}} \right) \right|}^{2}}{{\left| f\left( {{\omega }_{C}} \right) \right|}^{2}}{{e}^{i\left( {{\omega }_{A}}{{\tau }_{A}}+{{\omega }_{B}}{{\tau }_{B}}+{{\omega }_{C}}{{\tau }_{C}} \right)}}d{{\omega }_{A}}d{{\omega }_{B}}d{{\omega }_{C}}  
\end{aligned}
\label{10}\eea

Then we assume that each photon's frequency distributions $f\left( {{\omega }_{A}} \right)$, $f\left( {{\omega }_{B}} \right)$, $f\left( {{\omega }_{C}} \right)$ are Gaussian functions: 
\bea 
f\left( {{\omega }_{A}} \right)={{e}^{-\frac{{{\omega }_{A}}^{2}}{4\Delta \omega _{A}^{2}}}},f\left( {{\omega }_{B}} \right)={{e}^{-\frac{{{\omega }_{B}}^{2}}{4\Delta \omega _{B}^{2}}}},f\left( {{\omega }_{C}} \right)={{e}^{-\frac{{{\omega }_{C}}^{2}}{4\Delta \omega _{C}^{2}}}}
\label{11}\eea
where $\Delta {{\omega }_{A}}$,$\Delta {{\omega }_{B}}$,$\Delta {{\omega }_{C}}$ are the root mean square bandwidth of each photon and the central frequencies are set to 0. In this case, $R\left( {{\tau }_{A}},{{\tau }_{B}},{{\tau }_{C}} \right)$ may be calculated as
\bea  
R\left( {{\tau }_{A}},{{\tau }_{B}},{{\tau }_{C}} \right)\text{=exp}\left( -\frac{\Delta\omega _{A}^{2}\tau _{A}^{2}}{2}-\frac{\Delta\omega _{B}^{2}\tau _{B}^{2}}{2}-\frac{\Delta\omega _{C}^{2}\tau _{C}^{2}}{2} \right) 
\label{12}\eea

This result indicates that PMD leads to an exponential decay of the off-diagonal term, which is equivalent to a dephasing channel\cite{Siomau2010}.

For GHZ state undergoing PMD, the bipartite concurrence remains 0 and the expectation value of entanglement witness can be written as: 
\bea  V_{GHZ}^{1}\text{=}\frac{1}{2}-\frac{1\text{+}R\left( {{\tau }_{A}},{{\tau }_{B}},{{\tau }_{C}} \right)}{2}=-\frac{R\left( {{\tau }_{A}},{{\tau }_{B}},{{\tau }_{C}} \right)}{2} 
\label{13}\eea

In the case of W state, the concurrence between any two qubits becomes:

\bea C_{AB}^{1}=\frac{2R\left( {{\tau }_{A}},-{{\tau }_{B}},0 \right)}{3},C_{AC}^{1}=\frac{2R\left( {{\tau }_{A}},0,-{{\tau }_{C}} \right)}{3},C_{BC}^{1}=\frac{2R\left( 0,{{\tau }_{B}},-{{\tau }_{C}} \right)}{3}   
\label{14}\eea

And the expectation value of the entanglement witness is given by: 
\bea
\begin{aligned}
	V_{W}^{1}&\text{=}\frac{2}{3}-\frac{1+\frac{2}{3}R\left( {{\tau }_{A}},-{{\tau }_{B}},0 \right)+\frac{2}{3}R\left( {{\tau }_{A}},0,-{{\tau }_{C}} \right)+\frac{2}{3}R\left( 0,{{\tau }_{B}},-{{\tau }_{C}} \right)}{3}\\
	&=\frac{1-\frac{2}{3}R\left( {{\tau }_{A}},-{{\tau }_{B}},0 \right)-\frac{2}{3}R\left( {{\tau }_{A}},0,-{{\tau }_{C}} \right)-\frac{2}{3}R\left( 0,{{\tau }_{B}},-{{\tau }_{C}} \right)}{3}\\
\end{aligned}
\label{15}\eea

A relation between the expectation value of the tripartite entanglement witness and the bipartite concurrence may be inferred:
\bea
V_{W}^{1}\text{=}\frac{2}{3}-F=\frac{1-C_{AB}^{1}-C_{AC}^{1}-C_{BC}^{1}}{3}
\label{16}\eea
where $F\text{=}\frac{1+C_{AB}^{1}+C_{AC}^{1}+C_{BC}^{1}}{3}$ is the fidelity of the output state. The above relation shows that the entanglement decay rate of W states is equal to the average decay rate of three bipartite concurrences.

\subsection{Decoherence induced by PDL effects for GHZ and W states}
Next, we discuss the mode-filtering of the entangled states caused by the PDL effect. The quantum network consists of three fiber channels with PDL elements as illustrated in Fig. 2:
\begin{figure}[H]
	\centering
	\includegraphics[width=.75\columnwidth]{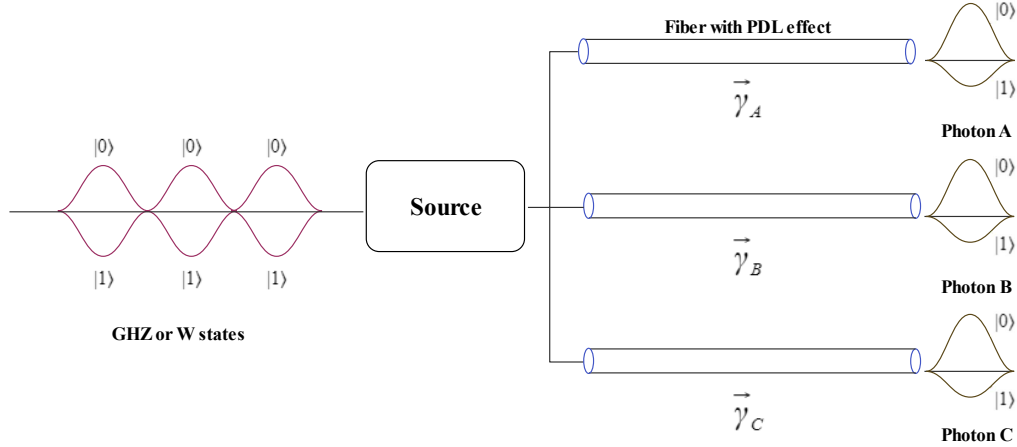}
	\caption{A quantum network with PDL magnitudes ${{\gamma }_{A}}$,${{\gamma }_{B}}$,${{\gamma }_{C}}$ where we assume that the polarization bases of each photon, $\left| 0 \right\rangle $ and $\left| 1 \right\rangle $, are aligned with the polarization states with maximum and minimum intensities in the channels. In this case, the polarization bases become the eigenstates of the operator$T$:$T\left| 0 \right\rangle \text{=}\exp \left( \frac{\gamma }{2} \right)\left| 0 \right\rangle T\left| 1 \right\rangle \text{=}\exp \left( \frac{-\gamma }{2} \right)\left| 1 \right\rangle $.} \label{Figure2}
\end{figure}

By applying the transmission matrix $T$ to the input polarization state, we can obtain the output polarization state:
\bea
\begin{aligned}
	 \psi _{GHZ}^{2}&=\exp \left( \frac{{{{\vec{\gamma }}}_{A}}\cdot \vec{\sigma }}{2} \right)\exp \left( \frac{{{{\vec{\gamma }}}_{B}}\cdot \vec{\sigma }}{2} \right)\exp \left( \frac{{{{\vec{\gamma }}}_{C}}\cdot \vec{\sigma }}{2} \right){{\psi }_{GHZ}} \\ 
	& =\frac{1}{\sqrt{2}}\left[ \exp \left( \frac{{{\gamma }_{A}}+{{\gamma }_{B}}+{{\gamma }_{C}}}{2} \right)\left| 000 \right\rangle \text{+}\exp \left( \frac{-{{\gamma }_{A}}-{{\gamma }_{B}}-{{\gamma }_{C}}}{2} \right)\left| 111 \right\rangle  \right], \\ 
	 \psi _{W}^{2}&=\exp \left( \frac{{{{\vec{\gamma }}}_{A}}\cdot \vec{\sigma }}{2} \right)\exp \left( \frac{{{{\vec{\gamma }}}_{B}}\cdot \vec{\sigma }}{2} \right)\exp \left( \frac{{{{\vec{\gamma }}}_{C}}\cdot \vec{\sigma }}{2} \right){{\psi }_{W}} \\ 
	& =\frac{1}{\sqrt{3}}\left[ \exp \left( \frac{{{\gamma }_{A}}+{{\gamma }_{B}}-{{\gamma }_{C}}}{2} \right)\left| 001 \right\rangle \text{+}\exp \left( \frac{{{\gamma }_{A}}-{{\gamma }_{B}}+{{\gamma }_{C}}}{2} \right)\left| 010 \right\rangle +\exp \left( \frac{-{{\gamma }_{A}}+{{\gamma }_{B}}+{{\gamma }_{C}}}{2} \right)\left| 100 \right\rangle  \right]  
\end{aligned}
\label{17}\eea
where ${{\gamma }_{A}}$,${{\gamma }_{B}}$,${{\gamma }_{C}}$ denote the loss coefficients in different channels. Then the density matrix of the GWZ states and W states after passing through the PDL elements takes the form:
\bea
\begin{aligned}
&{{\rho }_{GHZ}}=\frac{1}{2}\left( \begin{matrix}
	{{e}^{{{\gamma }_{A}}+{{\gamma }_{B}}+{{\gamma }_{C}}}} & 0 & 0 & 0 & 0 & 0 & 0 & 1  \\
	0 & 0 & 0 & 0 & 0 & 0 & 0 & 0  \\
	0 & 0 & 0 & 0 & 0 & 0 & 0 & 0  \\
	0 & 0 & 0 & 0 & 0 & 0 & 0 & 0  \\
	0 & 0 & 0 & 0 & 0 & 0 & 0 & 0  \\
	0 & 0 & 0 & 0 & 0 & 0 & 0 & 0  \\
	0 & 0 & 0 & 0 & 0 & 0 & 0 & 0  \\
	1 & 0 & 0 & 0 & 0 & 0 & 0 & {{e}^{-{{\gamma }_{A}}-{{\gamma }_{B}}-{{\gamma }_{C}}}}  \\
\end{matrix} \right),\\
&{{\rho }_{W}}=\frac{1}{3}\left( \begin{matrix}
	0 & 0 & 0 & 0 & 0 & 0 & 0 & 0  \\
	0 & {{e}^{{{\gamma }_{A}}+{{\gamma }_{B}}-{{\gamma }_{C}}}} & {{e}^{{{\gamma }_{A}}}} & 0 & {{e}^{{{\gamma }_{B}}}} & 0 & 0 & 0  \\
	0 & {{e}^{{{\gamma }_{A}}}} & {{e}^{{{\gamma }_{A}}+{{\gamma }_{C}}-{{\gamma }_{B}}}} & 0 & {{e}^{{{\gamma }_{C}}}} & 0 & 0 & 0  \\
	0 & 0 & 0 & 0 & 0 & 0 & 0 & 0  \\
	0 & {{e}^{{{\gamma }_{B}}}} & {{e}^{{{\gamma }_{C}}}} & 0 & {{e}^{{{\gamma }_{B}}+{{\gamma }_{C}}-{{\gamma }_{A}}}} & 0 & 0 & 0  \\
	0 & 0 & 0 & 0 & 0 & 0 & 0 & 0  \\
	0 & 0 & 0 & 0 & 0 & 0 & 0 & 0  \\
	0 & 0 & 0 & 0 & 0 & 0 & 0 & 0  \\
\end{matrix} \right)\\
\end{aligned}
\label{18}\eea

After normalization of the density matrces, the expectation value of the entanglement witness for GHZ state is given by
\bea 
V_{GHZ}^{2}\text{=}\frac{1}{2}-\left[ \frac{1}{2}\text{+}\frac{1}{2\cosh \left( {{\gamma }_{A}}+{{\gamma }_{B}}+{{\gamma }_{C}} \right)} \right]=-\frac{1}{2\cosh \left( {{\gamma }_{A}}+{{\gamma }_{B}}+{{\gamma }_{C}} \right)}
\label{19}\eea

For Bell states propagating in independent fiber channels, i.e. with different DGD ${{\tau }_{A}}$ and ${{\tau }_{B}}$, the concurrence reduces to $C=R\left( {{\tau }_{A}},{{\tau }_{B}} \right)\text{=exp}\left( -\frac{\Delta\omega _{A}^{2}\tau _{A}^{2}}{2}-\frac{\Delta\omega _{B}^{2}\tau _{B}^{2}}{2} \right)$\cite{Shtaif2011}. If only PDL effects are present with magnitudes ${{\gamma }_{A}}$ and ${{\gamma }_{B}}$, the concurrence is given by $C=\frac{1}{\cosh \left( {{\gamma }_{A}}+{{\gamma }_{B}} \right)}$\cite{Kirby2018}. From Eq.(13) and Eq.(19), we see that GHZ states show a similar decoherence behavior to the Bell state.

For the W state, the concurrence between any two qubits can be written as: 
\bea
\begin{aligned}
&C_{AB}^{2}=\frac{2{{e}^{{{\gamma }_{C}}}}}{{{e}^{{{\gamma }_{A}}+{{\gamma }_{B}}-{{\gamma }_{C}}}}\text{+}{{e}^{{{\gamma }_{A}}+{{\gamma }_{C}}-{{\gamma }_{B}}}}\text{+}{{e}^{{{\gamma }_{B}}+{{\gamma }_{C}}-{{\gamma }_{A}}}}},\\
&C_{AC}^{2}=\frac{2{{e}^{{{\gamma }_{B}}}}}{{{e}^{{{\gamma }_{A}}+{{\gamma }_{B}}-{{\gamma }_{C}}}}\text{+}{{e}^{{{\gamma }_{A}}+{{\gamma }_{C}}-{{\gamma }_{B}}}}\text{+}{{e}^{{{\gamma }_{B}}+{{\gamma }_{C}}-{{\gamma }_{A}}}}},\\
&C_{BC}^{2}=\frac{2{{e}^{{{\gamma }_{A}}}}}{{{e}^{{{\gamma }_{A}}+{{\gamma }_{B}}-{{\gamma }_{C}}}}\text{+}{{e}^{{{\gamma }_{A}}+{{\gamma }_{C}}-{{\gamma }_{B}}}}\text{+}{{e}^{{{\gamma }_{B}}+{{\gamma }_{C}}-{{\gamma }_{A}}}}}
\end{aligned}
\label{20}\eea
and the expectation value of the entanglement witness becomes: 
\bea
V_{W}^{2}\text{=}\frac{2}{3}-\left[ \frac{1}{3}\text{+}\frac{2\left( {{e}^{{{\gamma }_{A}}}}\text{+}{{e}^{{{\gamma }_{B}}}}\text{+}{{e}^{{{\gamma }_{C}}}} \right)}{3\left( {{e}^{{{\gamma }_{A}}+{{\gamma }_{B}}-{{\gamma }_{C}}}}\text{+}{{e}^{{{\gamma }_{A}}+{{\gamma }_{C}}-{{\gamma }_{B}}}}\text{+}{{e}^{{{\gamma }_{B}}+{{\gamma }_{C}}-{{\gamma }_{A}}}} \right)} \right]=\frac{1}{3}-\frac{2\left( {{e}^{{{\gamma }_{A}}}}\text{+}{{e}^{{{\gamma }_{B}}}}\text{+}{{e}^{{{\gamma }_{C}}}} \right)}{3\left( {{e}^{{{\gamma }_{A}}+{{\gamma }_{B}}-{{\gamma }_{C}}}}\text{+}{{e}^{{{\gamma }_{A}}+{{\gamma }_{C}}-{{\gamma }_{B}}}}\text{+}{{e}^{{{\gamma }_{B}}+{{\gamma }_{C}}-{{\gamma }_{A}}}} \right)}
\label{21}\eea

Also, in this case, a relation between the entanglement witness and the concurrence, similar to that expressed by Eq.(16), may be found
\bea
V_{W}^{2}\text{=}\frac{2}{3}-F=\frac{1-C_{AB}^{2}-C_{AC}^{2}-C_{BC}^{2}}{3}
\label{22}\eea

By comparing above results, we can draw the following conclusions: First, GHZ state decays in a form similar to that of the Bell state in fibers. Second, W states exhibit a universal relation between concurrence and the expectation value of entanglement witness. Finally, entanglement degradation for GHZ and W states differ for the same values of the noise parameters, indicating the different transmission performances of the two in a fiber network.

\section{Decoherence-free subspace and entanglement sudden death in fiber channels}
In this section, we discuss the conditions for ESD and DSF to occur when GHZ and W states propagate in fibers subject to PMD and PDL.
\subsection{Decoherence-free subspace of GHZ state}
In the previous section, we analyzed the decay of the entanglement in fiber channels without frequency correlations, which is similar to a dephasing channel. However, multi-photon entangled states prepared by SPDC sources are often characterized by frequency-correlated\cite{Wang2016}. For these states, which are hyperentangled in frequency and polarization, the evolution no longer leads to a simple exponential decay.  
Assuming that the three-photon entangled state is prepared by a continuous-wave(CW) pump\cite{Ruan2021}, the function $R\left( {{\tau }_{A}},{{\tau }_{B}},{{\tau }_{C}} \right)$ can be written as:
\bea
\begin{aligned}
	 R\left( {{\tau }_{A}},{{\tau }_{B}},{{\tau }_{C}} \right)&={{\iiint{\left| \tilde{f}\left( {{\omega }_{A}},{{\omega }_{B}},{{\omega }_{C}} \right) \right|}}^{2}}{{e}^{i\left( {{\omega }_{A}}{{\tau }_{A}}+{{\omega }_{B}}{{\tau }_{B}}+{{\omega }_{C}}{{\tau }_{C}} \right)}}d{{\omega }_{A}}d{{\omega }_{B}}d{{\omega }_{C}} \\ 
	& =\iiint{\left| f\left( \omega _A \right) \right|^2}\left| f\left( \omega _B \right) \right|^2\left| f\left( \omega _C \right) \right|^2\delta \left( \omega _A+\omega _B+\omega _C \right) e^{i\left( \omega _A\tau _A+\omega _B\tau _B+\omega _C\tau _C \right)}d\omega _Ad\omega _Bd\omega _C\\
\end{aligned}
\label{23}\eea
where $\delta \left( {{\omega }_{A}}\text{+}{{\omega }_{B}}\text{+}{{\omega }_{C}} \right)$ describes the frequency content of the pump lasers, and this kind of entangled states may be generated by cascaded spontaneous parametric downconversion\cite{Shalm2013}. With this assumption, $R\left( {{\tau }_{A}},{{\tau }_{B}},{{\tau }_{C}} \right)$ is given by:
\bea
R\left( {{\tau }_{A}},{{\tau }_{B}},{{\tau }_{C}} \right)\text{=exp}\left[ -\frac{\Delta\omega _{A}^{2}\Delta\omega _{B}^{2}{{\left( {{\tau }_{A}}-{{\tau }_{B}} \right)}^{2}}+\Delta\omega _{A}^{2}\Delta\omega _{C}^{2}{{\left( {{\tau }_{A}}-{{\tau }_{C}} \right)}^{2}}+\Delta\omega _{B}^{2}\Delta\omega _{C}^{2}{{\left( {{\tau }_{B}}-{{\tau }_{C}} \right)}^{2}}}{2\left( \Delta\omega _{A}^{2}+\Delta\omega _{B}^{2}+\Delta\omega _{C}^{2} \right)} \right]
\label{24}\eea
If the relation ${{\tau }_{A}}\text{=}{{\tau }_{B}}\text{=}{{\tau }_{C}}$ is satisfied, the expectation value of the entanglement witness of the output state is equivalent to the input state: 
\bea
V_{GHZ}^{1}\text{=}V_{GHZ}^{0}=-\frac{1}{2}
\label{25}\eea
i.e. GHZ states are immune to decoherence. The appearance of DSF means that as long as the magnitudes of the PMD in three channels are equal, the magnitude of the decoherence effect on each photon will be equal and cancel each other out. In this case, the photon output from each channel has the same time delay or time advance, which means it is not possible to track the information about polarization by detecting the arrival time of three photons.

DSF can also appear in channels affected by PDL. Since the value of the loss coefficient $\gamma $ is greater than zero, we can obtain:
\bea
V_{GHZ}^{2}\text{=}-\frac{1}{2\cosh \left( {{\gamma }_{A}}+{{\gamma }_{B}}+{{\gamma }_{C}} \right)}>-\frac{1}{2}
\label{26}\eea
But if we set the PDL vector ${{\vec{\gamma }}_{A}}$ in channel A to be anti-aligned to the PDL vectors ${{\vec{\gamma }}_{B}}$ and ${{\vec{\gamma }}_{C}}$ in channels B and C, the expectation value of the entanglement witness $V_{GHZ}^{2}$ becomes:
\bea
V_{GHZ}^{2}\text{=}-\frac{1}{2\cosh \left( -{{\gamma }_{A}}+{{\gamma }_{B}}+{{\gamma }_{C}} \right)}
\label{27}\eea
which implies that if the relation ${{\gamma }_{A}}\text{=}{{\gamma }_{B}}+{{\gamma }_{C}}$ is satisfied, DSF may be also obtained for GHZ state propagating in channels where PDL occurs: 
\bea
V_{GHZ}^{2}\text{=}V_{GHZ}^{0}=-\frac{1}{2}
\label{28}\eea

In addition, the expectation values of entanglement witness $V_{GHZ}^{1},V_{GHZ}^{2}<0$, and $V_{GHZ}^{1},V_{GHZ}^{2}$ are equal to zero only when the differential group delay ${{\tau }_{A}},{{\tau }_{B}},{{\tau }_{C}}$ or the loss coefficient ${{\gamma }_{A}},{{\gamma }_{B}},{{\gamma }_{C}}$ diverge. This suggests that no entanglement sudden death will occur for GHZ state.

The negative expectation value of the entanglement witness $-{{V}_{GHZ}}$ is shown in the following figures as a function of the differential group delay ${{\tau }_{A}}$ and loss coefficient ${{\gamma }_{A}}$, respectively:

\begin{figure}[H]
	\centering
	\includegraphics[width=.5\columnwidth]{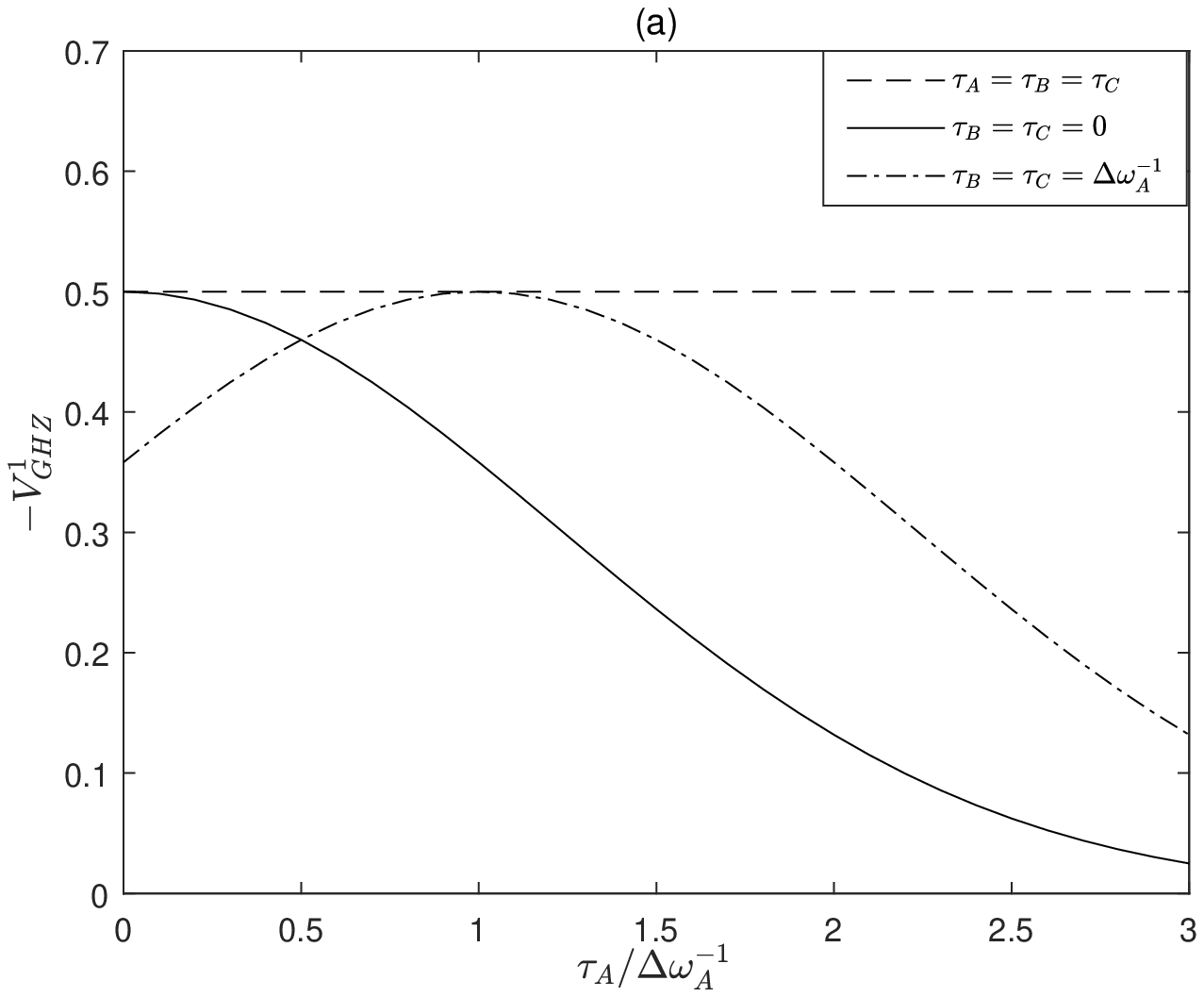}\hfill
	\includegraphics[width=.5\columnwidth]{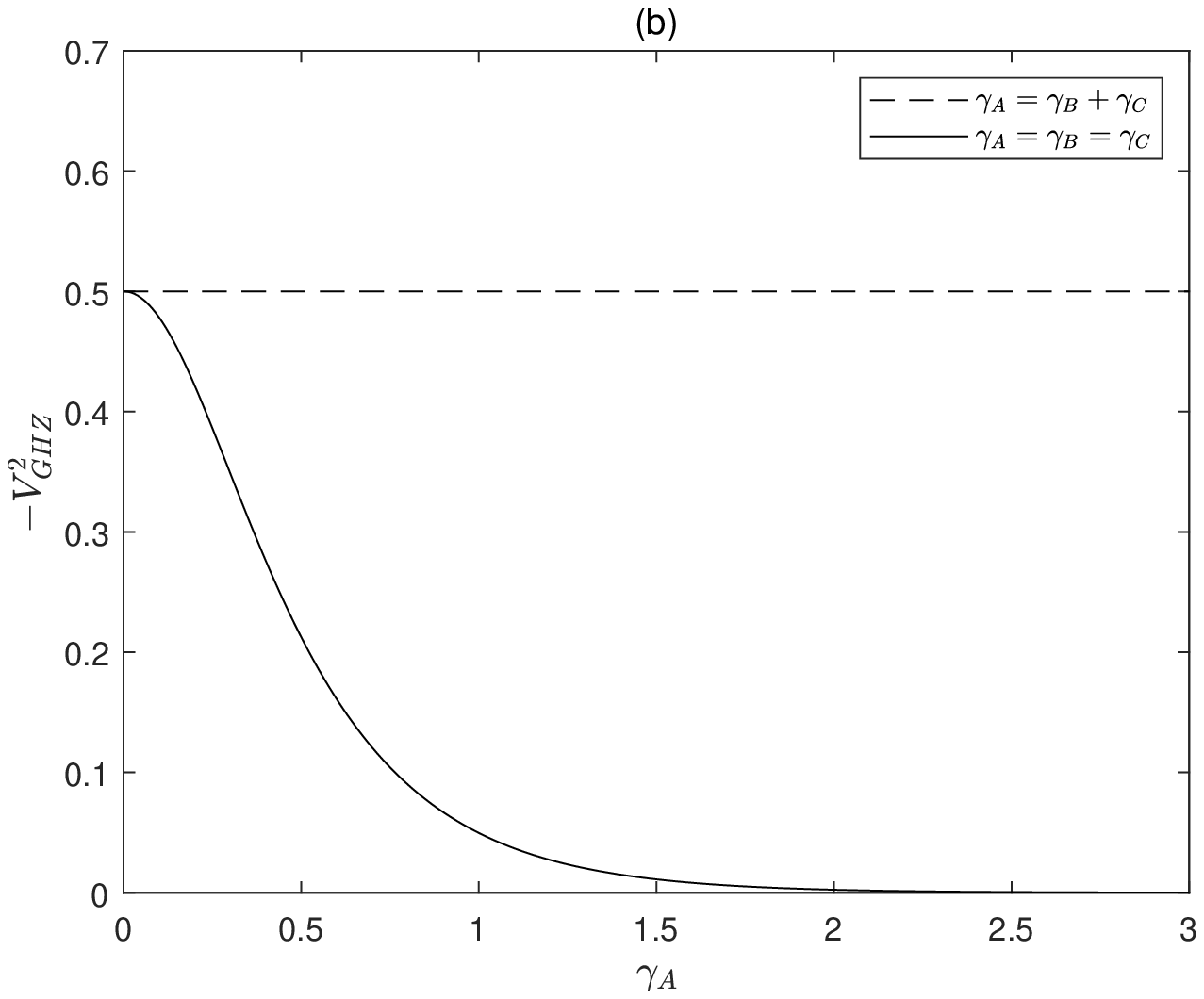}
	\caption{Decoherence behavior of GHZ states undergoing PMD and PDL effects. (a)The negative expectation value $-{{V}_{GHZ}^{1}}$ as a function of the DGD ${{\tau }_{A}}$ with $\Delta {{\omega }_{A}}=\Delta {{\omega }_{B}}=\Delta {{\omega }_{C}}$; The dashed line indicates the case where DSF occurs and the solid line indicates the monotonically decrease of entanglement with single-channel PMD, and the dash-dotted line shows the maximum entanglement achieved by the similar value of DGDs;  (b) The negative expectation value $-{{V}_{GHZ}^{2}}$ as a function of ${{\gamma }_{A}}$; The dashed line indicates the DSF induced by anti-aligned PDL vectors, and the solid line indicates the monotonically decrease of entanglement with aligned PDL vectors.}
\end{figure}
\subsection{Decoherence-free subspace and Entanglement sudden death of W state}
For W states propagating in fiber channels with PMD, when $R\left( {{\tau }_{A}},-{{\tau }_{B}},0 \right),R\left( {{\tau }_{A}},0,-{{\tau }_{C}} \right),R\left( 0,{{\tau }_{B}},-{{\tau }_{C}} \right)\le \frac{1}{2}$, we obtain ${{C}_{AB}},{{C}_{AC}},{{C}_{BC}}\le \frac{1}{3}$,$V_{W}^{1}\ge 0$. The magnitudes of PMD and PDL in the fiber increase with the photon’s transmission time and the fiber length. When the differential group delay accumulates to a certain extent, even if bipartite entanglement remains, tripartite entanglement may vanish in finite time, i.e. entanglement sudden death occurs.

For W states propagating in fibers with PDL, we can also have that when $\frac{{{e}^{{{\gamma }_{A}}}}\text{+}{{e}^{{{\gamma }_{B}}}}\text{+}{{e}^{{{\gamma }_{C}}}}}{{{e}^{{{\gamma }_{A}}+{{\gamma }_{B}}-{{\gamma }_{C}}}}\text{+}{{e}^{{{\gamma }_{A}}+{{\gamma }_{C}}-{{\gamma }_{B}}}}\text{+}{{e}^{{{\gamma }_{B}}+{{\gamma }_{C}}-{{\gamma }_{A}}}}}\le \frac{1}{2}$, $V_{W}^{2}\ge 0$. The above results mean that W states are more likely to experience ESD than GHZ states. An intuitive explanation of this phenomenon is that entanglement of W states is not inherently tripartite, but rather based on bipartite entanglement and, as such, more susceptible to decoherence.

For ${{\gamma }_{B}}\text{=}{{\gamma }_{C}}\text{=}0$, we can see that $\frac{{{e}^{{{\gamma }_{A}}}}\text{+}{{e}^{{{\gamma }_{B}}}}\text{+}{{e}^{{{\gamma }_{C}}}}}{{{e}^{{{\gamma }_{A}}+{{\gamma }_{B}}-{{\gamma }_{C}}}}\text{+}{{e}^{{{\gamma }_{A}}+{{\gamma }_{C}}-{{\gamma }_{B}}}}\text{+}{{e}^{{{\gamma }_{B}}+{{\gamma }_{C}}-{{\gamma }_{A}}}}}\text{=}\frac{{{e}^{{{\gamma }_{A}}}}\text{+}2}{2{{e}^{{{\gamma }_{A}}}}\text{+}{{e}^{-{{\gamma }_{A}}}}}>\frac{1}{2}$ and $V_{W}^{2}<0$. Remarkably, this result means that when PDL effect is only present in one fiber channel, the W state will never exhibit tripartite ESD. Furthermore, if we assume that ${{\gamma }_{A}}\to \infty $, we can get ${{C}_{AB}},{{C}_{AC}}\to 0,{{C}_{BC}}\to 1,V_{W}^{2}\to 0$, and $\psi _{W}^{2}$ in Eq.(17) can be changed to $\psi _{W}^{2}\to \frac{1}{\sqrt{2}}\left| 0 \right\rangle \left( \left| 01 \right\rangle \text{+}\left| 10 \right\rangle  \right)$. By performing mode-filtering operation on one photon, the W state can be decomposed into a Bell state and a separable photon.

Let us now address DSF for W states. Assuming that the three photons are frequency-correlated, we have:
\bea
R\left( {{\tau }_{A}},-{{\tau }_{B}},0 \right)\text{=exp}\left[ -\frac{\Delta\omega _{A}^{2}\Delta\omega _{B}^{2}{{\left( {{\tau }_{A}}+{{\tau }_{B}} \right)}^{2}}+\Delta\omega _{A}^{2}\Delta\omega _{C}^{2}{{\tau }_{A}}^{2}+\Delta\omega _{B}^{2}\Delta\omega _{C}^{2}{{\tau }_{B}}^{2}}{2\left( \Delta\omega _{A}^{2}+\Delta\omega _{B}^{2}+\Delta\omega _{C}^{2} \right)} \right]<1
\label{29}\eea

It is thus impossible for $V_{W}^{1}$ to be equal to $-\frac{1}{3}$ with non-zero DGD and no DSF for W states is expected in the presence of PMD. However, if the relation ${{\gamma }_{A}}\text{=}{{\gamma }_{B}}\text{=}{{\gamma }_{C}}$ is satisfied, the expectation value of the entanglement witness for W states undergoing PDL is degraded
\bea
V_{W}^{2}\text{=}V_{W}^{0}=-\frac{1}{3}
\label{30}\eea

This implies that W states under mode-filtering can also be immune to decoherence. In the following figures, we show the negative expectation value of the entanglement witness, illustrating the occurrence of decoherence-free subspace and entanglement sudden death for W states:
\begin{figure}[H]
	\centering
	\includegraphics[width=.5\columnwidth]{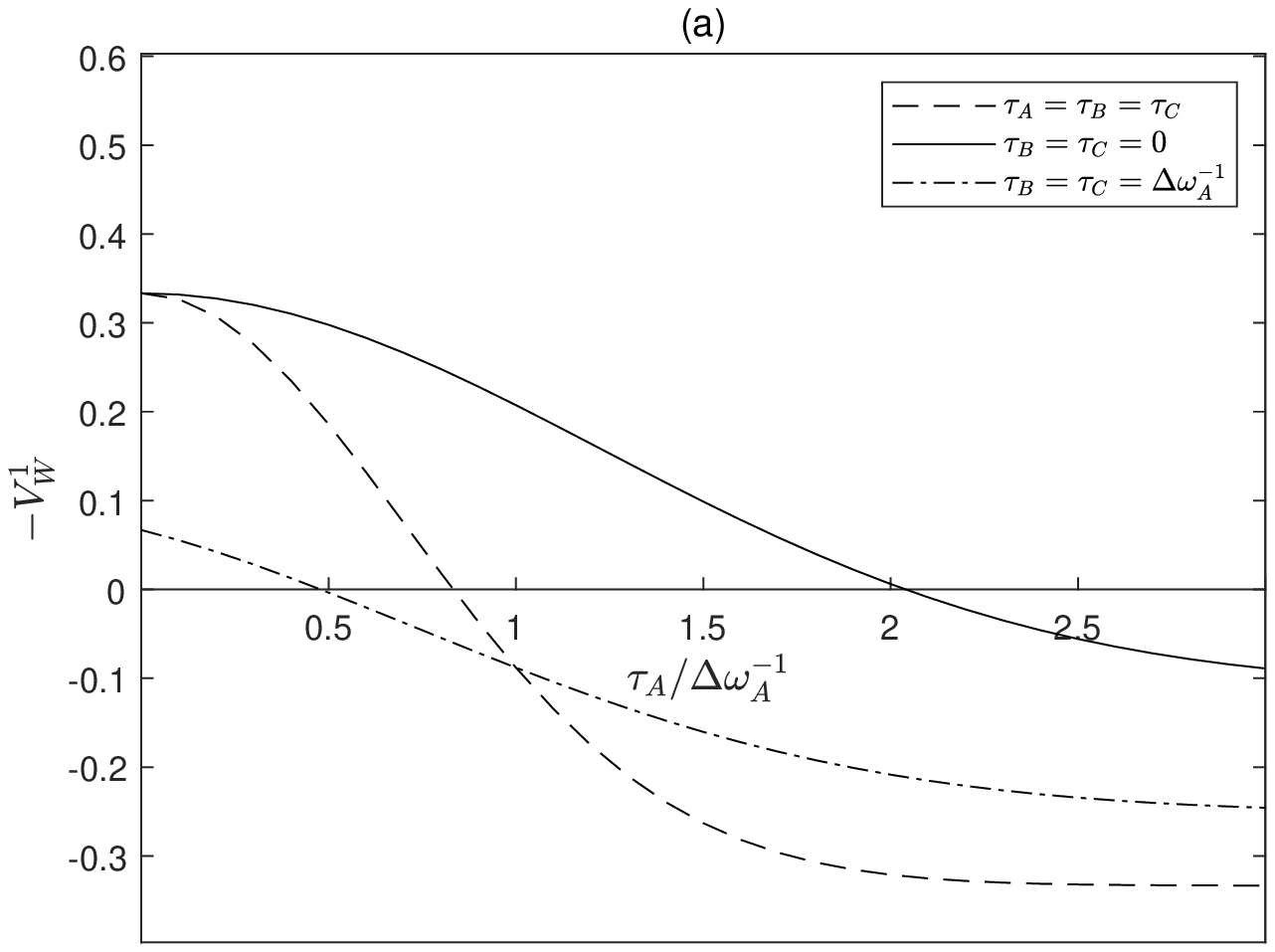}\hfill
	\includegraphics[width=.5\columnwidth]{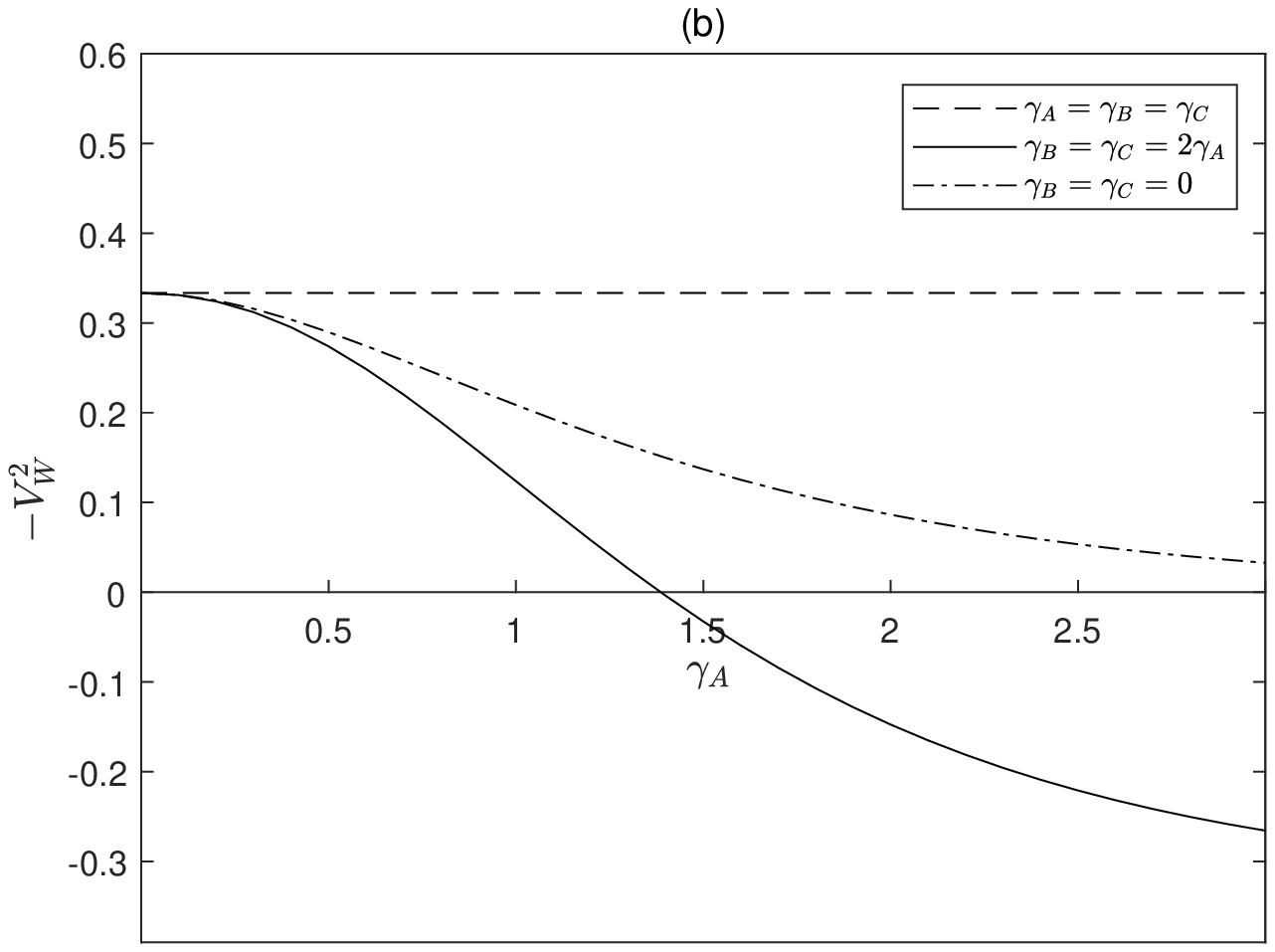}
	\caption{Decoherence behavior of W states undergoing PMD and PDL effects. (a)The negative expectation value $-{{V}_{W}^{1}}$ as a function of the DGD ${{\tau }_{A}}$ with $\Delta {{\omega }_{A}}=\Delta {{\omega }_{B}}=\Delta {{\omega }_{C}}$;  All three lines show that the W state entanglement decreases monotonically with the growing DGD and sudden death will occur; (b) The negative expectation value $-{{V}_{W}^{2}}$ as a function of ${{\gamma }_{A}}$; the dashed line indicates the DSF phenomenon with equal values of PDL and the solid line indicates the sudden death of entanglement and the dash-dotted line shows the W state does not experience ESD with single-channel PDL.}
\end{figure}

 In this Section, we have analyzed the appearance of decoherence-free subspace and the occurrence of entanglement sudden death for GHZ and W states propagating in fiber channels. We have found that GHZ states with frequency correlation can be immune to decoherence after passing through multiple fibers with the same PMD vectors. Similar DSF phenomena may also occur for both GHZ and W states in fibers exhibiting PDL. Unlike GHZ states, the tripartite ESD will occur for W states when the photons’ propagation time in the fiber exceeds a certain threshold.
\section{Decoherence of entangled states with an arbitrary number of photons in fiber channels}
  In this section, we try to extend the model from three to an arbitrary multiphoton entangled system. Here, we consider N-qubit GHZ and W states:
 \bea 
{{\left| {{\psi }_{N}} \right\rangle }_{GHZ}}\text{=}\frac{1}{\sqrt{2}}\left( \left| 00\cdots 0 \right\rangle \text{+}\left| 11\cdots 1 \right\rangle  \right),{{\left| {{\psi }_{N}} \right\rangle }_{W}}\text{=}\frac{1}{\sqrt{N}}\left( \left| 0\cdots 01 \right\rangle \text{+}\left| 0\cdots 10 \right\rangle +\left| 1\cdots 00 \right\rangle  \right)
\label{31}\eea

Each photon is numbered as 1,2,3,…, N. The concurrence of any two qubits is still 0 for GHZ states and is equal to 
$\frac{2}{N}$for W states. The expectation value of entanglement witness is $V_{GHZ}^{0}=Tr\left( EW{{\left| {{\psi }_{N}} \right\rangle }_{GHZ}}\left\langle  {{\psi }_{N}} \right| \right)=-\frac{1}{2}$ for $EW=\frac{1}{2}I-{{\left| {{\psi }_{N}} \right\rangle }_{GHZ}}\left\langle  {{\psi }_{N}} \right|$ and $V_{W}^{0}=Tr\left( EW{{\left| {{\psi }_{N}} \right\rangle }_{W}}\left\langle  {{\psi }_{N}} \right| \right)=-\frac{1}{N}$ for $EW=\frac{N-1}{N}I-{{\left| {{\psi }_{N}} \right\rangle }_{W}}\left\langle  {{\psi }_{N}} \right|$\cite{Sperling2013}.

 Assuming that each photon is transmitted individually through a fiber channel, we may repeat the calculations done in Section 3.3. The expectation values of the entanglement witness for N-qubit GHZ and W states undergoing PMD effects are given by:
 \bea 
\begin{aligned}
	& V_{GHZ}^{1}\text{=}\frac{1}{2}-\frac{1\text{+}R\left( {{\tau }_{1}},{{\tau }_{2}},\cdots ,{{\tau }_{N}} \right)}{2}=-\frac{R\left( {{\tau }_{1}},{{\tau }_{2}},\cdots ,{{\tau }_{N}} \right)}{2}, \\ 
	& V_{W}^{1}\text{=}\frac{N-1}{N}-F=\frac{N-2-\sum\limits_{j<k}^{N}{{{C}_{jk}}}}{N},j,k=1,2,\cdots ,N \\ 
\end{aligned}
\label{32}\eea
where ${{\tau }_{1}}$,${{\tau }_{2}}$,…,${{\tau }_{N}}$ denote the differential group delays in different channels, and ${{C}_{jk}}$ is the concurrence of any two photons in W states, i.e.
 \bea 
{{C}_{jk}}\text{=}\frac{2R\left( {{\tau }_{j}},-{{\tau }_{k}},0,\cdots ,0 \right)}{N}
\label{33}\eea

The function $R\left( {{\tau }_{1}},{{\tau }_{2}},\cdots ,{{\tau }_{N}} \right)$ is determined by the frequency distribution of the entangled photons:
 \bea 
R\left( {{\tau }_{1}},{{\tau }_{2}},\cdots ,{{\tau }_{N}} \right)={{\iiint{\left| \tilde{f}\left( {{\omega }_{1}},{{\omega }_{2}},\cdots ,{{\omega }_{N}} \right) \right|}}^{2}}{{e}^{i\left( {{\omega }_{1}}{{\tau }_{1}}+{{\omega }_{2}}{{\tau }_{2}}\cdots +{{\omega }_{N}}{{\tau }_{N}} \right)}}d{{\omega }_{1}}d{{\omega }_{2}}\cdots d{{\omega }_{N}}
\label{34}\eea

The multipartite entanglement of N-qubit GHZ and W states after passing through PDL elements can be characterized by:
 \bea 
\begin{aligned}
	& V_{GHZ}^{2}\text{=}\frac{1}{2}-\left[ \frac{1}{2}\text{+}\frac{1}{2\cosh \left( {{\gamma }_{1}}+{{\gamma }_{2}}\cdots +{{\gamma }_{N}} \right)} \right]=-\frac{1}{2\cosh \left( {{\gamma }_{1}}+{{\gamma }_{2}}\cdots +{{\gamma }_{N}} \right)}, \\ 
	& V_{W}^{2}\text{=}\frac{N-1}{N}-F=\frac{N-2-\sum\limits_{j<k}^{N}{{{C}_{jk}}}}{N},j,k=1,2,\cdots ,N \\ 
\end{aligned}
\label{35}\eea
where ${{\gamma }_{1}}$,${{\gamma }_{2}}$,…,${{\gamma }_{N}}$ denote the loss coefficients in different channels.${{C}_{jk}}$ can be written as a function of loss coefficients:
\bea
{{C}_{jk}}\text{=}{{e}^{-{{\gamma }_{j}}-{{\gamma }_{k}}}}\frac{2{{e}^{{{\gamma }_{1}}+{{\gamma }_{2}}\cdots \text{+}{{\gamma }_{N}}}}}{{{e}^{-{{\gamma }_{1}}+{{\gamma }_{2}}\cdots \text{+}{{\gamma }_{N}}}}\text{+}{{e}^{{{\gamma }_{1}}-{{\gamma }_{2}}\cdots \text{+}{{\gamma }_{N}}}}\cdots \text{+}{{e}^{{{\gamma }_{1}}+{{\gamma }_{2}}\cdots -{{\gamma }_{N}}}}}
\label{36}\eea

We can see that the entanglement dynamics for multi-particle GHZ states is similar to that of three-particle GHZ state and Bell states. For W states, we can express the expectation value of the entanglement witness in terms of the bipartite concurrences:
\bea
{{V}_{W}}\text{=}\frac{N-2-\sum\limits_{j<k}^{N}{{{C}_{jk}}}}{N}
\label{37}\eea

These results shows the amount of decoherence of a W state is related to the total amount of two-photon decoherence. For $0<\sum\limits_{j<k}^{N}{{{C}_{jk}}}\le N-2$, there is no detectable multipartite entanglement in the whole system even though there is still bipartite concurrence remaining. This relationship holds for the decoherence induced by both PMD and PDL and may not apply to other types of decoherence processes.
\section{Conclusion}
In conclusion, we have investigated the entanglement dynamics and decoherence mechanisms of multiphoton GHZ and W states propagating in fiber channels where PMD and PDL occur. By applying the transmission matrix to the input state, we obtain the density matrix, the concurrence, and the expectation value of specific entanglement witnesses at the output, and express them as functions of the DGD and the loss coefficient. We have found that in contrast to GHZ states, the entanglement of W states undergoing PMD and PDL vanishes in finite time, i.e. exhibits ESD. On the other hand, GHZ states may be, in certain conditions, immune to decoherence, even in the presence of PMD and PDL. If the PDL vectors in different fibers are equal, also W states may be made immune to decoherence induced by mode-filtering. Finally, we have discussed the decoherence process for W and GHZ states of an arbitrary number of qubits in fiber channels, and have found a relationship between the multipartite entanglement witness and the bipartite concurrence for W states.

Our study provides a detailed description of decoherence for multiphoton entangled states in fiber channels. Our results contribute to the characterization of multiqubit entanglement in noisy channels and provide a useful tool for the establishment of quantum networks. They also pave the way for further research, aimed at investigating the simultaneous decoherence induced by both PMD and PDL or involving entanglement in higher dimensions.

\section*{Acknowledgments}
This work was supported by National Key R$\&$D Program of China [grant numbers No. 2017YFE0301303]. We wish to express our gratitude to EditSprings for the expert linguistic services provided.

\end{document}